\documentclass[aps,prl,superscriptaddress,twocolumn,floatfix,longbibliography]{revtex4-2}  

\usepackage{graphicx,color}
\usepackage[colorlinks=true,linkcolor=blue,citecolor=blue,linkcolor=blue]{hyperref}%

\usepackage{amsmath,amssymb,amsfonts,bm}     
\usepackage{mathrsfs}
\usepackage{braket}
\allowdisplaybreaks[4]

\newcommand*{\dif}{\mathop{}\!\mathrm{d}}

\begin{document}
	
	
	\title{Chiral Casimir-Polder Force from Chiral Anomaly in Weyl semimetals}

	\author{Liang Chen}
	\affiliation{School of Mathematics and Physics, North China Electric Power University, Beijing 102206, China}
	\affiliation{Institute of Condensed Matter Physics, North China Electric Power University, Beijing 102206, China}
	
	\author{Xu-Feng Zhao}
	\affiliation{School of Mathematics and Physics, North China Electric Power University, Beijing 102206, China}
	\affiliation{Institute of Condensed Matter Physics, North China Electric Power University, Beijing 102206, China}

	\author{Xun-Lei Ding}
	\affiliation{School of Mathematics and Physics, North China Electric Power University, Beijing 102206, China}
	
	\author{Kai Chang}
	\email[Corresponding Email:]{kchang@zju.edu.cn}
	\affiliation{Center for Quantum Matter, School of Physics, Zhejiang University, Hangzhou 310027, China}
	\affiliation{Institute for Advanced Study in Physics, Zhejiang University, Hangzhou 310027, China}

\begin{abstract}
We propose a novel chiral-discriminatory Casimir-Polder interaction between chiral molecules and Weyl semimetals, originating from the chiral anomaly. This interaction is uniquely governed by the molecule's electric dipole moment, enabling chiral discrimination even when the optical rotatory strength of the molecule vanishes. For the chiral molecule C$_{20}$H$_{16}$ClNOS (characterized by strong spin-orbit coupling) and a rationally tailored Weyl semimetal, first-principles calculations predict a significantly enhanced chiral-discriminatory Casimir-Polder force compared to interactions between chiral molecule dimethyl disulphide and chiral metamaterial. We further propose a Stern-Gerlach-type spatial deflection measurement to isolate this chiral-specific interaction.
\end{abstract}

\maketitle

\emph{Introduction.---}The quantum vacuum in the vicinity of macroscopic bodies can be modulated by the quantum fluctuations of the electromagnetic (EM) field \cite{MilonniPW2013book}. This modulation offers a versatile framework for tailoring the intrinsic properties and dynamic behaviors of microscopic particles proximal to macroscopic bodies, with examples among them including Rabi splitting and spontaneous emission \cite{MorawitzH1974PRB,EagenCF1980CPL}, the Lamb shift \cite{Zubairybook,WangXH2004PRL,FuchsS2017PRA}, Casimir-Polder (CP) interaction \cite{Casimir1948PR2,SukenikCI1993PRL,HarberDM2005PRA,ObrechtJM2007PRL,BenderH2010PRL,GarcionC2021PRL,Bordag2009Casimirbook,SCheelBook,Buhmannbook,WesterbergN2023PhysRep}, van der Waals interactions between atoms \cite{SafariH2006PRA,SafariH2015PRL,BlockJ2017PRA,BarcellonaP2017PRL,SafariH2020NJP}, and the regulation of chemical reactions \cite{HutchisonJA2012ACIE,GalegoJ2019PRX,KeYZ2023PRL}. 
The seminal work \cite{Butcher2012NJP} establishes the chiral-discriminatory Casimir-Polder (CDCP) interaction by incorporating molecular chiral polarizability. 
This finding opens a new avenue for chiral separation. A Stern-Gerlach-type discriminator has been proposed to exploit this CDCP interaction \cite{SuzukiF2019PRA}. However, the contribution of the CP force arising from chiral polarizability is generally two orders of magnitude weaker than that induced by electric polarizability, as quantified by the relation $|\bm{\mu}/(c\bm{d})|\sim\alpha$. This CDCP force decreases further when the electric and magnetic dipole moments ($\bm{d}$ and $\bm{\mu}$) are non-parallel. 
For chiral plates composed of metamaterial, the molecule-plate separation must greatly exceed the microstructure's characteristic scale to avoid complex electrostatic interactions.
These constraints make CDCP force measurements highly challenging.

In this Letter, we propose a novel CDCP interaction between Pasteur chiral molecules and Weyl semimetals (WSMs) \cite{WanXG2011PRB,XuSY2015Science,LvBQ2015PRX,XuSY2015SciAdv}. This new CDCP interaction is induced by the chiral anomaly in WSMs. Chiral anomaly was first proposed in particle physics and developed to explain the decay process of the $\pi^0$ meson \cite{Adler1969PR,Bell1969INCA}. In WSMs, the chiral anomaly has distinctive physical effects, including the chiral magnetic effect \cite{LiQ2016NatPhys}, negative longitudinal magnetoresistance effect \cite{HuangCX2015PRX,ArnoldF2016NatComm,ReisRD2016NJP,LiangSH2018PRX}, non-local transport behavior \cite{ParameswaranSA2014PRX}, and anomalous thermoelectric property \cite{JiaZZ2016NatComm}. 
The unique EM response of WSMs suggests that the chiral-anomaly-induced Casimir-Lifshitz torque differs significantly from that in conventional birefringent materials \cite{ChenL2020PRL}. This motivates our study of CDCP interactions between chiral molecules and WSMs.

\begin{figure}[tb]
	\centering
	\includegraphics[width=\linewidth]{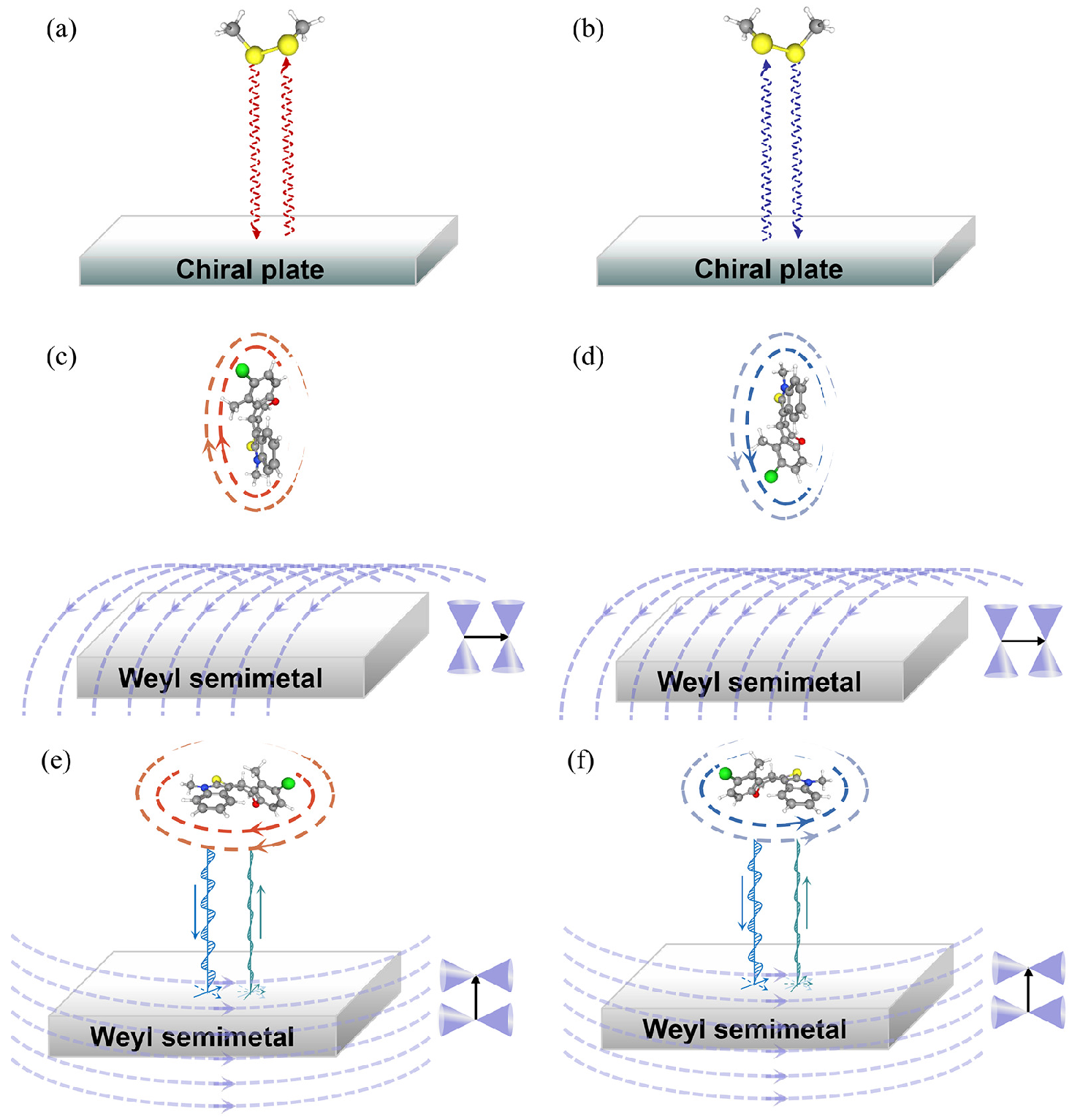}
	\caption{Schematic of the distinct mechanisms for the CDCP interaction. Molecules in the left and right columns are left- and right-handed, respectively. 
Top panel: CP interactions between the chiral molecule dimethyl disulphide (CH$_3$)$_2$S$_2$ and a chiral plate mediated by chiral photons, quantified by the optical rotatory strength.The red and blue wavy lines denote virtual left- and right-circularly polarized photons, respectively.
Middle and bottom panels: two distinct configurations of CP interactions between the chiral molecule C$_{20}$H$_{16}$ClNOS and a WSM, governed by the helicity of the virtual excited states in the molecule and characterized by the off-diagonal imaginary component of the electric dipole moment. Molecular orientations are specified for each configuration. Wavy lines in (e) and (f) represent possible virtual photon states between the molecule and the WSM; Kerr rotation of linearly polarized photon states is indicated by arrows on the WSM surface. 
The two light blue cones depict the linear dispersion of Weyl fermions within the Brillouin zone, with a black arrow marking the separation between Weyl nodes. Dashed circles surrounding the chiral molecules indicate virtual molecular excitations, while dashed arcs near the WSM represent quantum vacuum fluctuations arising from chiral anomaly. }
	\label{fig_1}
\end{figure}

The top panel in Fig. \ref{fig_1} depicts the conventional mechanism \cite{Butcher2012NJP}, wherein chiral molecules selectively interact with left- and right-circularly polarized light. 
The differential reflection of these photons is mediated by the chirality-dependent Fresnel coefficients of the surface, giving rise to a CDCP interaction.
This CDCP interaction scales linearly with the molecular optical rotatory strength, $R_{X,G}=\textrm{Im}[\bm{d}_{G,X}\cdot\bm{\mu}_{X,G}]$, where $\bm{d}_{G,X}$ and $\bm{\mu}_{X,G}$ denote the transition amplitudes of the electric and magnetic dipole moments, respectively. 
The middle and bottom panels in Fig. \ref{fig_1} illustrate two distinct configurations of the new mechanism. 
Driven by the chiral anomaly in WSMs, the quantum vacuum near the WSM surface exhibits quantum fluctuations of virtual helical photon states. For Pasteur chiral molecules with fixed orientations, enantiomers of opposite chirality undergo helical excitations of opposite handedness. The interaction between the virtual helical photon states and the helical molecular excitations make the CP interactions to be chiral discriminatory. This new CDCP interaction is independent of the molecular magnetic moment, instead, its chirality is essentially determined by the spin-orbit coupling (SOC) in the chiral molecule.

\emph{General theory.---}To provide a comprehensive investigation of this problem, we first derive a general formula for the CP interaction between a molecule and a plate. Under the dipole approximation, the molecule-EM field interaction Hamiltonian takes the form, 
\begin{gather}
\hat{H}=\hat{H}_{\mathrm{M}}+\hat{H}_{\mathrm{int}}+\hat{H}_{\mathrm{EM}}, \\
\hat{H}_{\mathrm{int}}=\int\left[\delta(\bm{r}-\bm{r}_0)\left(\hat{\bm{E}}\cdot\hat{\bm{d}}+\hat{\bm{B}}\cdot\hat{\bm{\mu}}\right)\right]{\dif}\bm{r}, \\
\hat{H}_{\mathrm{EM}}=\frac{1}{2}\int\left(\varepsilon_0\hat{\bm{E}}^2+\frac{1}{\mu_0}\hat{\bm{B}}^2\right){\dif}\bm{r}, 
\end{gather}
and $\hat{H}_{\mathrm{M}}=\sum_{X}E_{X}\ket{X}\bra{X}$ describes the molecular internal dynamics, $\hat{\bm{d}}$ and $\hat{\bm{\mu}}$ are the electric and magnetic dipole moment operators, $\hat{\bm{E}}$ and $\hat{\bm{B}}$ represent the electric and magnetic field operators, $\bm{r}_0=(0,0,a)$ defines the location of the chiral molecule, the plate surface is located at $z=0$. Under the Coulomb gauge ($\hat{\bm{E}}=-\partial_t{\hat{\bm{A}}}$, $\hat{\bm{B}}=\bm{\nabla}\times\hat{\bm{A}}$), the EM field can be quantized as
\begin{gather}
\hat{\bm{A}}(\bm{r},t)= \sum_{\lambda}  
\int\sqrt{\frac{\hbar}{2\omega_{\bm{k}}}} 
\left[\hat{A}_{\lambda}(\bm{k})\bm{\varphi}_{\lambda}(\bm{k},\bm{r})+h.c.\right]{\dif}\bm{k}, 
\end{gather}
where $\omega_{\bm{k}}=ck$ (with $c$ the speed of light) is the quantized eigenfrequency, $\hat{A}_{\lambda}(\bm{k})$ is the annihilation operator for the EM mode with wavevector $\bm{k}$ and polarization $\lambda$ ($\lambda=\textrm{TE}$ and $\textrm{TM}$ refer to the transverse electric and transverse magnetic modes, respectively), $h.c.$ denotes the Hermitian conjugate. The mode function $\bm{\varphi}_{\lambda}(\bm{k},\bm{r})$ is determined by the reflection and transmission coefficients of EM waves on the plate surface (see the Supplemental Material \cite{supplemental} for details). 
Employing the standard quantum field theory \cite{Negelebook}, we derive the zero-temperature CP potential:   
\begin{gather}
\mathcal{E}_{\mathrm{CP}}(a)=\frac{1}{\varepsilon_0}\sum_{X>G}\int\frac{d\bm{k}_{\parallel}}{(2\pi)^2}\int_{-\infty}^{0}\frac{dk_z}{2\pi}\textrm{Re}\bigg\{\frac{\omega_{\bm{k}}e^{-2ik_za}}{\omega_{X,G}+\omega_{\bm{k}}} \notag\\
\textrm{Tr}[\mathsf{D}_{X,G}\cdot\mathsf{S}(\bm{k}_{\parallel},k_z)]\bigg\}. \label{eqX5}
\end{gather}
Here $\bm{k}=(\bm{k}_{\parallel},k_z)$ is the photon wavevector,
$\bm{k}_{\parallel}=(k_x,k_y)$ its in-plane projection, and
$\omega_{\bm{k}}=c|\bm{k}|$.  The sum runs over excited molecular
states $\ket{X}$ at energies $\hbar\omega_{X,G}=E_X-E_G$ above
the ground state $\ket{G}$.
The $6\times6$ tensor $\mathsf{D}_{X,G}$ collects the electric and
magnetic dipole matrix elements,
\begin{equation}
\mathsf{D}_{X,G}=\left(\begin{matrix}
\bm{d}_{X,G}\bm{d}_{G,X} & \bm{d}_{X,G}\bm{\mu}_{G,X} \\
\bm{\mu}_{X,G}\bm{d}_{G,X} & \bm{\mu}_{X,G}\bm{\mu}_{G,X}
\end{matrix}\right),  \nonumber 
\end{equation}
with $\bm{d}_{X,G}=\bra{X}\hat{\bm{d}}\ket{G}$,
$\bm{\mu}_{X,G}=\bra{X}\hat{\bm{\mu}}\ket{G}$.
The scattering matrix $\mathsf{S}(\bm{k}_{\parallel},k_z)$ encodes the
EM wave reflection on the surface:
\begin{gather}
\mathsf{S}(\bm{k}_{\parallel},k_z)=\sum_{\lambda,\lambda^{\prime}} r_{\lambda,\lambda^{\prime}} \left[\begin{matrix}
\bm{\epsilon}_{\lambda^{\prime}}(\bm{k}_R)\bm{\epsilon}_{\lambda}(\bm{k})  & \bm{\epsilon}_{\lambda^{\prime}}(\bm{k}_R)\tilde{\bm{\epsilon}}_{\lambda}(\bm{k})/c \\
\tilde{\bm{\epsilon}}_{\lambda^{\prime}}(\bm{k}_R)\bm{\epsilon}_{\lambda}(\bm{k})/c  & \tilde{\bm{\epsilon}}_{\lambda^{\prime}}(\bm{k}_R)\tilde{\bm{\epsilon}}_{\lambda}(\bm{k})/c^2
\end{matrix}\right],  \nonumber
\end{gather}
where $r_{\lambda,\lambda^{\prime}}$ ($\lambda,\lambda^{\prime}=\mathrm{TE},\mathrm{TM}$)
are the elements of the $2\times2$ Fresnel matrix,
$\bm{k}_R=(\bm{k}_{\parallel},-k_z)$ is the specularly reflected
wavevector, $\bm{\epsilon}_{\lambda}(\bm{k})$ is the polarization
vector of the electric field, and
$\tilde{\bm{\epsilon}}_{\mathrm{TE}}=\bm{\epsilon}_{\mathrm{TM}}$,
$\tilde{\bm{\epsilon}}_{\mathrm{TM}}=-\bm{\epsilon}_{\mathrm{TE}}$
are the magnetic polarization vectors.
A complete derivation of this result is included in the Supplemental Material \cite{supplemental}. 
Rotating the $k_z$-integral in Eq. (\ref{eqX5}) to the imaginary axis ($\omega_{\bm{k}}\to i\zeta$) yields the standard imaginary-frequency representation of CP potential \cite{Bordag2009Casimirbook,SCheelBook,Buhmannbook,WesterbergN2023PhysRep}: 
\begin{gather}
\mathcal{E}_{\mathrm{CP}}(a)=\frac{1}{\varepsilon_0c^2}\sum_{X>G}\int\frac{{\dif}\bm{k}_{\parallel}}{(2\pi)^2}\int_0^{\infty}\frac{{\dif}\zeta}{2\pi}\frac{e^{-2\kappa{a}}}{\kappa} \notag \\
\mathrm{Re}\left\{\frac{\zeta^2}{\omega_{X,G}+i\zeta}\mathrm{Tr}\left[\mathsf{D}_{X,G}\cdot\mathsf{S}(\bm{k}_{\parallel},-i\kappa)\right]\right\},
\label{eqX8}
\end{gather}
where $\zeta$ is the imaginary frequency, $\kappa=\sqrt{k_{\parallel}^2+\zeta^2/c^2}$. It encompasses symmetric \cite{Casimir1948PR2,Bordag2009Casimirbook,SCheelBook,Buhmannbook,WesterbergN2023PhysRep} and asymmetric \cite{MarachevskyVN2010PRD,BuhmannSY2018PRA,MarachevskyVN2023PRD} electric polarizabilities, magnetic polarizability \cite{BoyerTH1969PR,SafariH2008PRA,BimonteG2009PRA}, and chiral susceptibility \cite{Butcher2012NJP}.  In this work, we focus on the $\mathsf{S}$-matrix of a WSM surface, whose permittivity tensor, modified by the chiral anomaly, enables the purely electric channel to mediate a chiral-discriminatory CP interaction. We first illustrate the general framework using two representative cases,  and then analyze the new mechanisms depicted in Fig. \ref{fig_1}.

\emph{Case 1.---}For isotropic molecules with only the electric dipole contribution, Eq. (\ref{eqX8}) reduces to the standard zero-temperature Casimir-Polder potential \cite{Bordag2009Casimirbook}, 
\begin{gather}
\mathcal{E}_{\mathrm{CP}}^{(d)}(a)=-\frac{1}{\varepsilon_0}\int_0^{\infty}\frac{{k}_{\parallel}{\dif}{k}_{\parallel}}{(2\pi)}\int_0^{\infty}\frac{{\dif}\zeta}{2\pi}\alpha(i\zeta)\kappa{e^{-2\kappa{a}}} \notag \\ 
\left[2r_{\textrm{TM,TM}}-\frac{\zeta^2}{c^2\kappa^2}\left(r_{\textrm{TE,TE}}+r_{\textrm{TM,TM}}\right)\right], \label{eq_X11}
\end{gather}
where $\alpha(i\zeta)$ is the molecular dynamic polarizability. 

\emph{Case 2.---}For isotropic chiral molecules with solely the chiral contribution, the dipole matrix tensor reduces to the form, 
\begin{equation}
\mathsf{D}_{X,G}=\frac{iR_{X,G}}{3}\left(\begin{matrix}
0 & \mathsf{I} \\
-\mathsf{I} & 0
\end{matrix}\right), 
\end{equation}
and we obtain, 
\begin{gather}
\mathcal{E}_{\mathrm{CP}}^{(R)}(a)=\frac{1}{3\varepsilon_0}\sum_{X>G}\frac{R_{X,G}}{c}\int\frac{{\dif}\bm{k}_{\parallel}}{(2\pi)^2}\int_0^{\infty}\frac{{\dif}\zeta}{2\pi}\frac{e^{-2\kappa{a}}}{\kappa} \notag \\ 
\textrm{Re}\left[\frac{2ik_{\parallel}^2}{\omega_{X,G}+i\zeta}(r_{\textrm{TE,TM}}-r_{\textrm{TM,TE}})\right]. \label{eq_X13}
\end{gather}
When $r_{\textrm{TE,TM}} = -r_{\textrm{TM,TE}}$, Eq. (\ref{eq_X13}) reduces to the non-resonant part of zero-temperature CP potential given in Ref. \cite{Butcher2012NJP}.

\emph{Case 3.---}We next consider the configurations sketched in the middle panel of Fig. \ref{fig_1}, focusing on the scenario where chiral discrimination arises exclusively from the electric dipole terms. To this end, we employ a two-level model of molecule to illustrate the underlying physics,  
\begin{equation}
\hat{H}_{\mathrm{M}}=E_G\ket{G}\bra{G}+E_X\ket{X}\bra{X}, 
\end{equation}
where the ground state $\ket{G}$ has angular momentum quantum numbers $(l,m_z)=(0,0)$; for enantiomers of opposite chiralities, the excited state $\ket{X}$ exhibits opposite magnetic quantum numbers: $(l,m_y)=(1,{\pm}1)$ for right-handed (RH) and left-handed (LH) molecules, respectively. The electric dipole matrix tensor, $\bm{d}_{G,X}\bm{d}_{X,G}$, can be written as, 
\begin{equation}
\bm{d}_{G,X}\bm{d}_{X,G}=\left(\begin{matrix}
0 & 0 & -i\chi_{X,G} \\
0 & 0 & 0 \\
i\chi_{X,G} & 0  & 0
\end{matrix}\right), 
\end{equation}
where $\chi_{X,G}=\pm|\int{r^3}\psi_{1,0}^*(r)\psi_{2,1}(r)dr|^2/6$ (the $\pm$ sign distinguishes RH and LH molecules), and $\psi_{n,l}(r)$ denotes the radial wave function with principal quantum number $n$ and angular quantum number $l$. Within this two-level framework, the zero-temperature CDCP potential arising from electric dipole terms is derived as:  
\begin{gather}
\mathcal{E}_{\mathrm{CP}}^{(\chi)}(a)=\frac{1}{\varepsilon_0}\sum_{X>G}\chi_{X,G}\int\frac{{\dif}\bm{k}_{\parallel}}{(2\pi)^2}\int_0^{\infty}\frac{{\dif}\zeta}{2\pi}\frac{e^{-2\kappa{a}}}{\kappa} \mathrm{Re}\bigg\{
\notag \\
\frac{\zeta^2}{\omega_{X,G}+i\zeta}\left[\frac{2\kappa{k_x}}{\zeta^2}r_{\textrm{TM,TM}}+\frac{k_y}{c\zeta}\left(r_{\textrm{TM,TE}}-r_{\textrm{TE,TM}}\right)\right]\bigg\}. \label{eq_X17}
\end{gather}
A non-zero $\mathcal{E}_{\mathrm{CP}}^{(\chi)}(a)$ requires the reflection coefficients in the imaginary-frequency representation to exhibit odd symmetry in $\bm{k}_{\parallel}$. This condition is inherently fulfilled by WSMs due to their intrinsic chiral anomaly. Specifically, when the Weyl node separation in the Brillouin zone is oriented along the $x$-direction, the  imaginary parts of the Fresnel coefficients exhibit the following odd-symmetry relations \cite{ChenL2020PRL}, 
\begin{gather}
\textrm{Im}[r_{\textrm{TE,TM}}(k_x,-k_y)]=-\textrm{Im}[r_{\textrm{TE,TM}}(k_x,k_y)], \\
\textrm{Im}[r_{\textrm{TE,TM}}(-k_x,k_y)]=+\textrm{Im}[r_{\textrm{TE,TM}}(k_x,k_y)], \\
\textrm{Im}[r_{\textrm{TM,TM}}(k_x,-k_y)]=+\textrm{Im}[r_{\textrm{TM,TM}}(k_x,k_y)], \\
\textrm{Im}[r_{\textrm{TM,TM}}(-k_x,k_y)]=-\textrm{Im}[r_{\textrm{TM,TM}}(k_x,k_y)]. 
\end{gather}
Notably, the real components of these coefficients retain even symmetry with respect to $\bm{k}_{\parallel}$, which does not contribute to chiral discrimination.

\begin{figure}[tb]
	\centering
	\includegraphics[width=\linewidth]{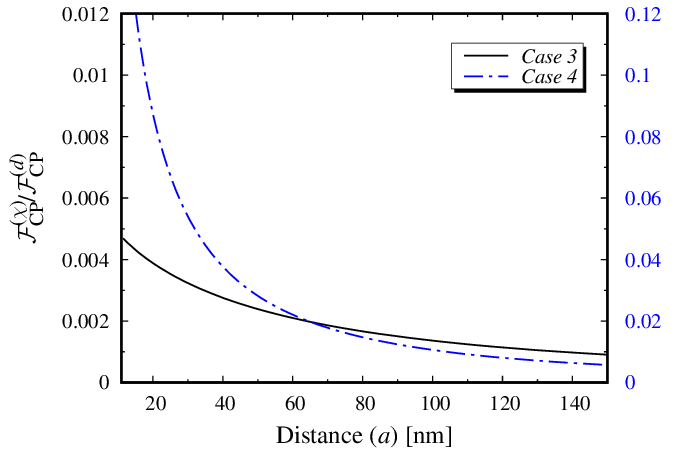}
	\caption{Relative strengths of CDCP force as a function of distance between the molecule C$_{20}$H$_{16}$ClNOS and WSM plate for configurations given in \emph{Case 3} (Black solid line, left axis) and \emph{Case 4} (Blue dashed line, right axis).}
	\label{fig_2}
\end{figure}

Physically, the imaginary part of $d_{G,X}^{x}d_{X,G}^z$ ($\chi_{X,G}$) encodes the helicity of the chiral molecule. This helicity signature originates from the molecular SOC. Therefore, the new mechanism should be most prominent in molecules with strong SOC, where such helicity-dependent effects are amplified. We illustrate this with a practical example: the organic molecule
19-chloro-9,20-dimethyl-15-oxa-11-thia-9-azapentacyclo[11.8.0.0$^{2,10}$.0$^{3,8}$.0$^{16,21}$]henicosa-2(10),3,5,7,12,16(21),17,19-octaene \cite{pubchem_cid102166221}, which features a three-dimensional helical configuration and pronounced SOC. For brevity, we refer to this molecule by its molecular formula, 
C$_{20}$H$_{16}$ClNOS, 
in subsequent discussions. We calculate the molecular electronic structure using the quantum chemistry package ORCA 6.0 \cite{NeeseF2025WCMS} employing the B3LYP functional \cite{BeckeAD1993JCP,LeeC1988PRB,VoskoSH1980CJP,StephensPJ1994JPC} and def2-TZVP basis set \cite{WeigendF2005PCCP}. 
Excited-state properties are further investigated via time-dependent density functional theory, with explicit inclusion of SOC to capture fine-structure effects and compute transition dipole matrix elements. We find: the average electric dipole polarizability $\sum_{j=x,y,z}d_{G,X}^jd_{X,G}^j/3=4.133\times10^{-2}$ au$^2$, the transition frequency $\omega_{X,G}=4.6618$ eV$/\hbar$, and $\chi_{X,G}=1.528\times10^{-4}$ au$^2$. For the WSM, we derive its optical conductivity and permittivity using a low-energy effective model, with Weyl node separation oriented along the $x$-direction and set to $0.8$ nm$^{-1}$. The Fresnel matrix is computed via the transfer matrix method (Details on the WSM permittivity functions and the derivation of the Fresnel matrix can be found in the Supplementary Material \cite{supplemental}).    The black solid line in Fig. \ref{fig_2} shows the relative strength of CDCP force, defined as $\mathcal{F}_{\textrm{CP}}^{(\chi)}(a)/\mathcal{F}_{\textrm{CP}}^{(d)}(a)$, where
\begin{equation}
\mathcal{F}^{(c)}_{\textrm{CP}}(a)=-\frac{\partial\mathcal{E}^{(c)}_{\textrm{CP}}(a)}{\partial{a}},  ~~~ (c=\chi, d),
\end{equation}
$\mathcal{E}^{(\chi)}_{\textrm{CP}}(a)$ and $\mathcal{E}^{(d)}_{\textrm{CP}}(a)$ are given in Eqs. (\ref{eq_X17}) and (\ref{eq_X13}), respectively. 
For separations $20$ nm $<a<$ $100$ nm, this ratio is on the order of $10^{-3}$ ---  
a magnitude consistents with prior calculations of CDCP interaction based on optical rotatory strength in a distinct system (dimethyl disulfide in a chiral metamaterial cavity \cite{Butcher2012NJP}). 

\emph{Case 4.---}The CDCP interaction can be enhanced by adopting an alternative configuration (the bottom panel in Fig. \ref{fig_1}): the separation of the two Weyl nodes is oriented perpendicular to the WSM surface, the chiral molecule is rotated to a corresponding orientation, with its excited-state wavefunction given by $\langle{\bm{r}}|X\rangle=\psi_{2,1}(r)Y_{1,1}(\vartheta,\varphi)$. For this configuration, the electric dipole matrix tensor becomes, 
\begin{equation}
\bm{d}_{G,X}\bm{d}_{X,G}=\left(\begin{matrix}
0 & i\chi_{X,G} & 0 \\
-i\chi_{X,G} & 0 & 0 \\
0& 0  & 0
\end{matrix}\right). \label{eq_X22}
\end{equation}
Substituting this result into the general formula, Eq. (\ref{eqX8}), yields the CDCP potential for this configuration: 
\begin{gather}
\mathcal{E}_{\mathrm{CP}}^{(\chi)}(a)=\frac{1}{\varepsilon_0}\sum_{X>G}\chi_{X,G}\int\frac{{\dif}\bm{k}_{\parallel}}{(2\pi)^2}\int_0^{\infty}\frac{{\dif}\zeta}{2\pi}\frac{e^{-2\kappa{a}}}{\kappa} 
\notag \\
\mathrm{Re}\left\{\frac{\zeta^2}{\omega_{X,G}+i\zeta}\left[i\frac{\kappa}{c\zeta}\left(r_{\textrm{TM,TE}}+r_{\textrm{TE,TM}}\right)\right]\right\}. \label{eq_X23}
\end{gather}

\begin{figure}[tb]
	\centering
	\includegraphics[width=\linewidth]{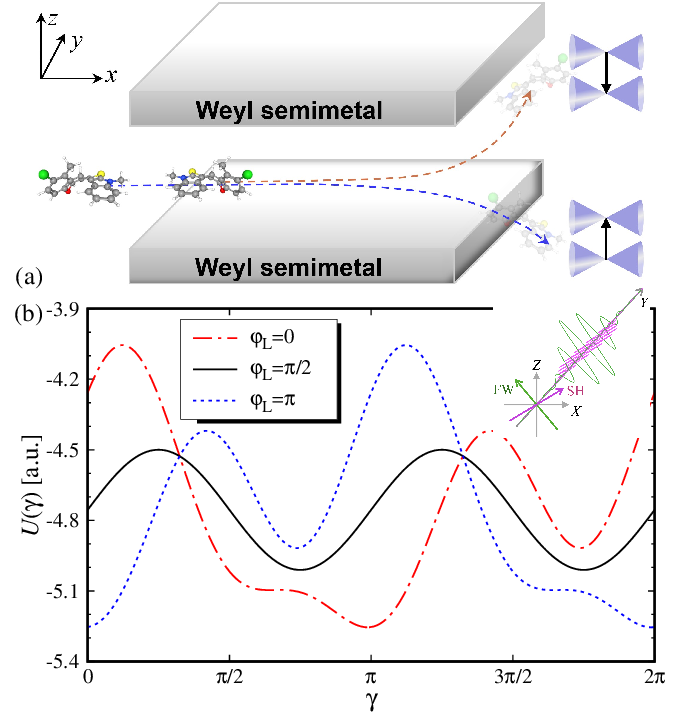}
	\caption{ (a) Schematic of the Stern-Gerlach-type experiment to detect the CDCP interaction. Two WSM plates are arranged in parallel. The separations of Weyl nodes are aligned perpendicular to the plate surfaces. (b) Potential energy $U(\gamma)$ of the C$_{20}$H$_{16}$ClNOS molecule under a phase-locked cross-polarized femtosecond laser pulse. Here $\gamma$ denotes the molecular orientation angle in the $xz$-plane, and $\varphi_L$ represents the relative phase between the FW and SH fields. The insert illustrates the configuration of the synchronized two-color laser pulses. }
	\label{fig_3}
\end{figure}

Although the CDCP potentials in Eqs. (\ref{eq_X13}) and (\ref{eq_X23}) both rely on the off-diagonal components of the Fresnel matrix, their origins are distinct. Equation (\ref{eq_X13}) originates from the differential reflection of circularly polarized photons on the chiral plate surface, as manifested by the relation $r_{\textrm{TE,TM}}-r_{\textrm{TM,TE}}=i(r_{++}-r_{--})$ in the chiral basis \cite{JiangQD2019PRB}. Equation (\ref{eq_X23}), however, stems from optical nonreciprocity (Kerr rotation illustrated in Fig. \ref{fig_1}) at the WSM surface, where the circular dichroism cancels in $r_{\textrm{TM,TE}}+r_{\textrm{TE,TM}}$. 
The CP interaction involving the asymmetric components of the polarizability in the form of Eq. (\ref{eq_X22}), combined with a Chern-Simons-type boundary condition that induces the nonreciprocal response $r_{\textrm{TM,TE}}=r_{\textrm{TE,TM}}$, has been studied in related works \cite{MarachevskyVN2010PRD,BuhmannSY2018PRA,MarachevskyVN2023PRD}. To isolate the effect of this nonreciprocity, we consider a limiting case: if we replace the WSM with a perfectly reflecting nonreciprocal mirror where $r_{\textrm{TM,TE}}=r_{\textrm{TE,TM}}=1$, we obtain the analytical result: 
\begin{equation}
\mathcal{E}_{\mathrm{CP}}^{(\chi)}(a)=\sum_{X>G}\frac{\chi_{X,G}}{8\pi^2\varepsilon_0{a^3}}\left[1-\frac{\tilde{a}}{\sqrt{\pi}}\mathscr{G}_{1,3}^{3,1}\left(\tilde{a}^2\Bigg|\begin{matrix} \frac{1}{2} \\ 0,\frac{1}{2},\frac{3}{2} \end{matrix} \right) \right], 
\end{equation}
where $\mathscr{G}_{1,3}^{3,1}$ is the Meijer-G function, $\tilde{a}=\omega_{X,G}a/c$ is the dimensionless distance. The asymptotic behavior of the Meijer-G function  yields the same short- and long-distance scaling as found for Chern–Simons‑type boundary conditions \cite{MarachevskyVN2010PRD,BuhmannSY2018PRA}. Specifically, in the small-distance limit ($\tilde{a}\ll1$), $\mathcal{E}_{\mathrm{CP}}^{(\chi)}(a)=\sum_{X>G}{\chi_{X,G}}/({8\pi^2\varepsilon_0{a^3}})$ (dominated by the $a^{-3}$ term); in the large-distance limit ($\tilde{a}\gg1$), $\mathcal{E}_{\mathrm{CP}}^{(\chi)}(a)=\sum_{X>G}{\chi_{X,G}}/({8\pi^2\varepsilon_0\omega_{X,G}^2{a^5}})$ (transitioning to an $a^{-5}$ dependence). For comparison, we revisit the electric dipole-dominated CP potential, Eq. (\ref{eq_X11}), under ideal metal boundary conditions ($r_{\textrm{TM,TM}}=-r_{\textrm{TE,TE}}=1$). As shown in Ref. \cite{Bordag2009Casimirbook}, $\mathcal{E}_{\mathrm{CP}}^{(d)}(a)=-\sum_{X>G}{\bm{d}_{G,X}\cdot\bm{d}_{X,G}}/({48\pi\varepsilon_0{a^3}})$ for $a\rightarrow0$, and $\mathcal{E}_{\mathrm{CP}}^{(d)}(a)=-\sum_{X>G}{\bm{d}_{G,X}\cdot\bm{d}_{X,G}}/({16\pi^2\varepsilon_0\omega_{X,G}{a^4}})$ for $a\rightarrow\infty$. The ratio of CP forces $\mathcal{F}_{\textrm{CP}}^{(\chi)}(a)/\mathcal{F}_{\textrm{CP}}^{(d)}(a)$ tends to a constant in the short distance limit and be proportional to $a^{-1}$ in the large distance limit. For the C$_{20}$H$_{16}$ClNOS-WSM system investigated here, numerical calculations reveal a more complex dependence due to the WSM's chiral anomaly-induced electromagnetic response: in the short distance limit ($a\rightarrow0$), we find $\mathcal{F}_{\textrm{CP}}^{(\chi)}(a)/\mathcal{F}_{\textrm{CP}}^{(d)}(a)\propto{a^{-1}}$, and in the large distance limit ($a\rightarrow1~\mu\textrm{m}$), $\mathcal{F}_{\textrm{CP}}^{(\chi)}(a)/\mathcal{F}_{\textrm{CP}}^{(d)}(a)\propto{a^{-3/2}}$. Notably, 
numerical results show that this ratio reaches $10\%$ at $a=20$ nm and remains on the order of $1\%$ for $a{\lesssim}100$ nm (see Fig. \ref{fig_2}, blue dashed line).

Fig. \ref{fig_3}(a) shows a proposed Stern-Gerlach-type setup to detect the CDCP force.  Two WSM plates are placed parallel. The separation of their Weyl nodes in the Brillouin zone aligned perpendicular to the surfaces (along the $z$-direction). This can be achieved by selecting specific crystal surfaces: for WSM candidates EuCd$_2$As$_2$ \cite{MaJZ2019SciAdv,WangLL2019PRB,SohJR2019PRB}, KCrTe and RbCrTe \cite{LiuHS2024PRB}, we can choose both the upper and lower surfaces to be (001). The resulting CDCP forces from the two plates are complementary and chiral distinct. The WSM plates separation is $100$ nm, chiral molecules are injected into the gap from the left side, with an initial distance of $50$ nm to each WSM plate, such that the achiral contributions to the CP force are mutually cancelled. Enantiomers of C$_{20}$H$_{16}$ClNOS can be spatially separated by the CDCP force. Quantitative calculations reveal that the magnitude of this new CDCP force is approximately $2\times10^{-21}$ N. The acceleration experienced by the molecules is estimated to be on the order of $10^3$ m/$\textrm{s}^2$. To manipulate the molecular orientation, we leverage femtosecond laser pulses, a state-of-the-art technique for molecular orientation control \cite{LinK2018NatComm}. Specifically, we propose using a short phase-locked orthogonal two-color pulse polarized within the $xz$-plane. 
The polarization of the fundamental wave (FW) is tilted at $+45^{\circ}$ from the $z$-axis, and correspondingly, the polarization of the second harmonic (SH) is tilted at $+45^{\circ}$ from the $x$-axis, as illustrated in the inset of Fig. \ref{fig_3}.
For the intensities, $I_{\textrm{FW}}=2.2\times10^{15}\textrm{W}/\textrm{cm}^2$ and $I_{\textrm{SH}}=4.7\times10^{14}\textrm{W}/\textrm{cm}^2$, we calculate the time-averaged potential energy of C$_{20}$H$_{16}$ClNOS molecule in this field as a function of its orientation angle $\gamma$ in the $xz$-plane (see Fig. \ref{fig_3}(b) for results and Supplemental Material \cite{supplemental} for detailed calculations). For the relative phase $\varphi_L=\pi$, the potential energy exhibits a global minimal at $\gamma=0$, the molecule is oriented in the $z$-direction.

\emph{Remarks.---}Two points warrant further clarification. First, the two-plate cavity (Fig.~\ref{fig_3}) introduces interference corrections beyond the sum of two independent single-surface CP forces. For the achiral CP force, symmetry guarantees exact cancellation at the cavity center even with cavity corrections included. Away from the center, the leading correction scales as the square of the reflection coefficients; the CDCP cavity correction follows the same scaling. Numerical estimates put both corrections at the percent level, rendering them negligible. A complete derivation of the cavity corrections is provided in the Supplemental Material \cite{supplemental}. Second, we do not include the contribution from Fermi arc surface states. These states, while a consequence of bulk-boundary correspondence, do not affect the bulk EM responses driven by the chiral anomaly. Hence, Fermi arcs generate no chiral-discriminatory force of their own, and the predicted CDCP signal is robust against their presence.

\emph{Summary.---}In this Letter, we propose a novel mechanism for CDCP interaction by exploiting the intrinsic helicity of chiral molecules and the chiral anomaly in WSMs. By conducting quantitative analysis of the chiral molecule C$_{20}$H$_{16}$ClNOS and WSM materials, we establish that the CDCP interaction arising from this new mechanism is observable for molecules centered between the WSM plates and precisely oriented by femtosecond laser pulses. This mechanism establishes a pioneering approach for manipulating chiral molecules in vacuum quantum engineering, especially in scenarios where the conventional CDCP interactions are ineffective. 
Our theoretical framework exhibits generalizability to other topological quantum systems, such as topological insulators, quantum anomalous Hall systems, and Chern insulators.

\emph{Acknowledgments.---}LC gratefully acknowledges Lu Liu for insightful discussions on femtosecond laser technology. 
This work was supported by NSFC under Grant Nos. 12488101 and 12174101. 
LC and XFZ are partially supported by the Fundamental Research Funds for the Central Universities under Grant No. 2026SL013.


%

\end{document}